\documentclass[twocolumn,prl,aps,superscriptaddress]{revtex4}

\usepackage{graphicx}
\usepackage{dcolumn}
\begin{document}

\title{Observation of Two-source Interference in
the Photoproduction Reaction $Au Au
\rightarrow Au Au \rho^0$}
 
\affiliation{Argonne National Laboratory, Argonne, Illinois 60439, USA}
\affiliation{University of Birmingham, Birmingham, United Kingdom}
\affiliation{Brookhaven National Laboratory, Upton, New York 11973, USA}
\affiliation{University of California, Berkeley, California 94720, USA}
\affiliation{University of California, Davis, California 95616, USA}
\affiliation{University of California, Los Angeles, California 90095, USA}
\affiliation{Universidade Estadual de Campinas, Sao Paulo, Brazil}
\affiliation{Carnegie Mellon University, Pittsburgh, Pennsylvania 15213, USA}
\affiliation{University of Illinois at Chicago, Chicago, Illinois 60607, USA}
\affiliation{Creighton University, Omaha, Nebraska 68178, USA}
\affiliation{Nuclear Physics Institute AS CR, 250 68 \v{R}e\v{z}/Prague, Czech Republic}
\affiliation{Laboratory for High Energy (JINR), Dubna, Russia}
\affiliation{Particle Physics Laboratory (JINR), Dubna, Russia}
\affiliation{Institute of Physics, Bhubaneswar 751005, India}
\affiliation{Indian Institute of Technology, Mumbai, India}
\affiliation{Indiana University, Bloomington, Indiana 47408, USA}
\affiliation{Institut de Recherches Subatomiques, Strasbourg, France}
\affiliation{University of Jammu, Jammu 180001, India}
\affiliation{Kent State University, Kent, Ohio 44242, USA}
\affiliation{University of Kentucky, Lexington, Kentucky, 40506-0055, USA}
\affiliation{Institute of Modern Physics, Lanzhou, China}
\affiliation{Lawrence Berkeley National Laboratory, Berkeley, California 94720, USA}
\affiliation{Massachusetts Institute of Technology, Cambridge, MA 02139-4307, USA}
\affiliation{Max-Planck-Institut f\"ur Physik, Munich, Germany}
\affiliation{Michigan State University, East Lansing, Michigan 48824, USA}
\affiliation{Moscow Engineering Physics Institute, Moscow Russia}
\affiliation{City College of New York, New York City, New York 10031, USA}
\affiliation{NIKHEF and Utrecht University, Amsterdam, The Netherlands}
\affiliation{Ohio State University, Columbus, Ohio 43210, USA}
\affiliation{Old Dominion University, Norfolk, VA, 23529, USA}
\affiliation{Panjab University, Chandigarh 160014, India}
\affiliation{Pennsylvania State University, University Park, Pennsylvania 16802, USA}
\affiliation{Institute of High Energy Physics, Protvino, Russia}
\affiliation{Purdue University, West Lafayette, Indiana 47907, USA}
\affiliation{Pusan National University, Pusan, Republic of Korea}
\affiliation{University of Rajasthan, Jaipur 302004, India}
\affiliation{Rice University, Houston, Texas 77251, USA}
\affiliation{Universidade de Sao Paulo, Sao Paulo, Brazil}
\affiliation{University of Science \& Technology of China, Hefei 230026, China}
\affiliation{Shanghai Institute of Applied Physics, Shanghai 201800, China}
\affiliation{SUBATECH, Nantes, France}
\affiliation{Texas A\&M University, College Station, Texas 77843, USA}
\affiliation{University of Texas, Austin, Texas 78712, USA}
\affiliation{Tsinghua University, Beijing 100084, China}
\affiliation{United States Naval Academy, Annapolis, MD 21402, USA}
\affiliation{Valparaiso University, Valparaiso, Indiana 46383, USA}
\affiliation{Variable Energy Cyclotron Centre, Kolkata 700064, India}
\affiliation{Warsaw University of Technology, Warsaw, Poland}
\affiliation{University of Washington, Seattle, Washington 98195, USA}
\affiliation{Wayne State University, Detroit, Michigan 48201, USA}
\affiliation{Institute of Particle Physics, CCNU (HZNU), Wuhan 430079, China}
\affiliation{Yale University, New Haven, Connecticut 06520, USA}
\affiliation{University of Zagreb, Zagreb, HR-10002, Croatia}

\author{B.~I.~Abelev}\affiliation{University of Illinois at Chicago, Chicago, Illinois 60607, USA}
\author{M.~M.~Aggarwal}\affiliation{Panjab University, Chandigarh 160014, India}
\author{Z.~Ahammed}\affiliation{Variable Energy Cyclotron Centre, Kolkata 700064, India}
\author{B.~D.~Anderson}\affiliation{Kent State University, Kent, Ohio 44242, USA}
\author{D.~Arkhipkin}\affiliation{Particle Physics Laboratory (JINR), Dubna, Russia}
\author{G.~S.~Averichev}\affiliation{Laboratory for High Energy (JINR), Dubna, Russia}
\author{Y.~Bai}\affiliation{NIKHEF and Utrecht University, Amsterdam, The Netherlands}
\author{J.~Balewski}\affiliation{Massachusetts Institute of Technology, Cambridge, MA 02139-4307, USA}
\author{O.~Barannikova}\affiliation{University of Illinois at Chicago, Chicago, Illinois 60607, USA}
\author{L.~S.~Barnby}\affiliation{University of Birmingham, Birmingham, United Kingdom}
% ...

\collaboration{STAR Collaboration}\noaffiliation

\begin{abstract}


In ultra-peripheral relativistic heavy-ion collisions, a photon from the
electromagnetic field of one nucleus can fluctuate to a
quark-antiquark pair and scatter from the other nucleus, emerging as a
$\rho^0$.  The $\rho^0$ production occurs in two well-separated
(median impact parameters of 20 and 40 fermi for the cases considered here)
nuclei, so the system forms
a 2-source interferometer.  At low transverse momenta, the two
amplitudes interfere destructively, suppressing $\rho^0$  production.  
Since the $\rho^0$ decays before the production
amplitudes from the two sources can overlap, the two-pion system can only be
described with an entangled non-local wave function, and is thus an example
of the Einstein-Podolsky-Rosen paradox. 
We observe this suppression in 200 GeV per nucleon-pair gold-gold collisions.
The interference is $87\!\% \!\pm 5\!\% {\rm (stat.)}\!\pm 8\!\%$ (syst.) of the 
expected level.  This translates into a
limit on decoherence due to wave function collapse or other
factors, of 23\% at the 90\% confidence level.

\end{abstract}

\maketitle

\narrowtext

Relativistic heavy ions carry strong electromagnetic fields which can 
be treated as sources of quasi-real virtual photons. When two ions collide,
a large variety of two-photon and
photonuclear interactions can occur \cite{reviews}.  In
coherent vector meson photoproduction, a photon from the field of one
nucleus fluctuates into a virtual quark-antiquark pair which scatters
elastically from the other nucleus, emerging as a real vector meson.  $\rho^0$ photoproduction
has a large cross section, 8-10\% of the hadronic cross
section for gold-gold collisions at a center-of-mass energy of 200 GeV
per nucleon-pair \cite{usPRC,STARrho^0,strikman}.  Photoproduction can occur at large impact
parameters, $b$. For $\rho^0$ photoproduction the median $b$ is about 46 fm \cite{baltzus}. 

The $(q\overline q)N$ scattering that produces $\rho^0$ occurs via the short-ranged strong force;
the $\rho^0$ is produced within one of the two ions.  The $\rho^0$ source consists of two
well-separated nuclei.  There are two possibilities:
either nucleus 1 emits a photon which
scatters off nucleus 2, or vice versa.  These two possibilities are
indistinguishable, and are related by a parity transformation. Vector mesons have
negative parity, so the two amplitudes combine with opposite signs.
The nuclear separation can be accounted for with
a transverse momentum ($p_T$) dependent phase factor.  The cross section is \cite{interfere}
\begin{equation}
\sigma(p_T,b,y)\! =\! \bigg| A(p_T,b,y) - A(p_T,b,-y)\exp{(i\vec{p}_T\cdot \vec{b})}\bigg|^2,
\label{eq:sigmay}
\end{equation}
where $A(p_T,b,y)$ and $A(p_T, b,-y)$ are the amplitudes at rapidity $y$ for $\rho^0$ production from
the two photon directions.  We take $\hbar=c=1$ here. 
At mid-rapidity the amplitudes for the two directions are equal, and
\begin{equation}
\sigma(p_T,b,0) = 2|A(p_T,b,0)|^2 \bigg[1-\cos{(\vec{p}_T\cdot \vec{b})}\bigg].
\end{equation}
The
system acts as a 2-slit interferometer, with slit separation $b=|\vec{b}|$.  
The cross-sections at different $\vec{b}$ are added, and
the $p_T$ spectrum is obtained by integrating Eq. (1) over $\vec{b}$.  $\rho^0$ production is
suppressed for $p_T\!\lesssim\! 1/\langle b\rangle$,
where $\langle b\rangle$ is the mean impact parameter.  

The $\rho^0$ rapidity $y$ and mass $m_V$ and the
photon energy $k_i$ are related by
$k_{1,2} = (m_V/2) \exp(\pm y)$ where the subscript refers to the two
directions.  Away from $y=0$, $k_1\ne k_2$, so $A(p_T,b,y)\ne A(p_T,b,-y)$, 
and the interference in Eq. (\ref{eq:sigmay}) is less than maximal. 

There are two theoretical calculations of this interference.  Klein
and Nystrand \cite{interfere} calculated the interference using a detailed nuclear 
form factor, averaging the photon flux over the nucleus.
Hencken, Baur and Trautmann used
a more detailed model of the photon profile and a
Gaussian form factor for the nucleus \cite{BHT}.  This work only
considered production at mid-rapidity ($y=0$), and so cannot be directly compared 
with the data presented here.   At $y=0$, the two calculations agree quite well.

If the $\rho^0$ production phase depends on the photon energy, this would
introduce a $y-$dependent phase shift into Eq. (\ref{eq:sigmay}).
This is not expected in the soft-Pomeron model \cite{phase}, and we assume that 
this phase difference is negligible.

The produced $\rho^0$s decay almost immediately at two well-separated points, so
any interference must develop after the decay, and involve the $\pi^+\pi^-$ final state.
Since the pions go in different directions, this requires an entangled $\pi^+\pi^-$ wave function
which cannot be factorized into separate $\pi^+$ and $\pi^-$ wave functions;
this is an example of the Einstein-Podolsky-Rosen paradox \cite{physlett,EPR}.  A measurement of
the two-source interference is sensitive to any loss of quantum mechanical coherence,
be it due to interactions with the environment \cite{decoherencereview} or as 
a characteristic of the $\rho^0$ decay.

Interference is also expected when the $\rho^0$ photoproduction is accompanied by
mutual Coulomb excitation of the two nuclei.  This reaction proceeds primarily via 
three independent single-photon subreactions 
(one to excite each nucleus, and one to produce the $\rho^0$) \cite{baltzus}.
At a given $b$, the cross-section
for the subreactions factorizes; the probability for an
$n$-photon reaction is $P_n(b) = \prod_{i=1}^n P_i(b)$ \cite{factorize},
where $P_i(b)$ is the probability for subreaction $i$.
Therefore, these  multi-photon reactions have much smaller $\langle b \rangle$ and
the effect of interference extends to higher $p_T$ \cite{baltzus}.  
Because of the different $\langle b\rangle$,
multi-photon interactions are important for studying this interference.
The Klein-Nystrand model uses measured photonuclear cross-sections for
the mutual Coulomb excitation \cite{bcw},
while Hencken, Baur and Trautmann used the Giant Dipole Resonance, plus a correction.  
For $e^+e^-$ production accompanied by nuclear breakup, using the measured mutual
breakup cross-sections rather than the Hencken  Baur and Trautmann approach
leads to a 20\% larger cross-section \cite{newbaltz}; a similar difference may apply
for $\rho^0$ photoproduction. 

In this letter we measure two-source interference in 200 GeV per nucleon-pair
gold-gold collisions by studying the transverse momentum ($p_T$) spectrum
of photoproduced $\rho^0$s.  These data were taken with the STAR detector.
The major detector component used here is a central Time Projection 
Chamber (TPC) \cite{TPC} in a 0.5 T solenoidal magnet.  The TPC tracked charged particles with 
pseudorapidity $|\eta|<1.0$.  We used two trigger
detector systems, the Central Trigger Barrel (CTB) and two Zero Degree Calorimeters
(ZDCs).  The CTB consisted of
240 scintillator slats surrounding the TPC, detecting
charged particles with pseudorapidity $|\eta|<1.0$ \cite{trigger}.  The 
ZDCs detected neutrons emitted by the dissociating gold nuclei with virtually
unchanged longitudinal momentum (100 GeV/$c$) and low $p_T$ \cite{ZDCs}.

Data were collected with two different trigger conditions.  The first
was a topology trigger which selected events with
roughly back-to-back pions in the CTB \cite{STARrho^0}.  Nearly vertical pairs 
were excluded, to reduce contamination from cosmic rays.
The second, minimum bias (MB) trigger selected $\rho^0$ accompanied by
mutual dissociation.  In these events, both nuclei broke up and released 
neutrons into the two ZDCs. The cross-section for $\rho^0$ production accompanied by
mutual Coulomb excitation is about 7\% \cite{baltzus} of the total $\rho$ photoproduction
cross-section, so the two datasets are essentially independent. 

Events were required to have net charge zero and 
exactly two reconstructed tracks which formed a vertex less than 50 cm 
longitudinally from the center  of the TPC for
the MB sample, and 100 cm for the topology sample.  The difference is because
events from the CTB based trigger were distributed more broadly along this axis. 
For the topology data, we exclude
events with $|y|<0.05$ to remove the remaining contamination 
from cosmic rays, where a single muon track could be reconstructed as 
two tracks with net charge 0, $p_T=0$ and $y=0$.
All tracks were assumed to be pions, and were required to have a $\pi\pi$
invariant mass 550 MeV/$c^2$ $< M_{\pi\pi} < 920$ MeV/$c^2$. These criteria
produced a clean set of $\rho^0$ events, at some
cost in efficiency.  The 550 MeV/$c^2$ mass cut removes background from 
misidentified two-photon production of lepton pairs.
The background, estimated from the like-sign 
pion pairs, was small, 1.4\%.

\begin{figure}[hbtp]
\includegraphics[clip,scale=0.42]{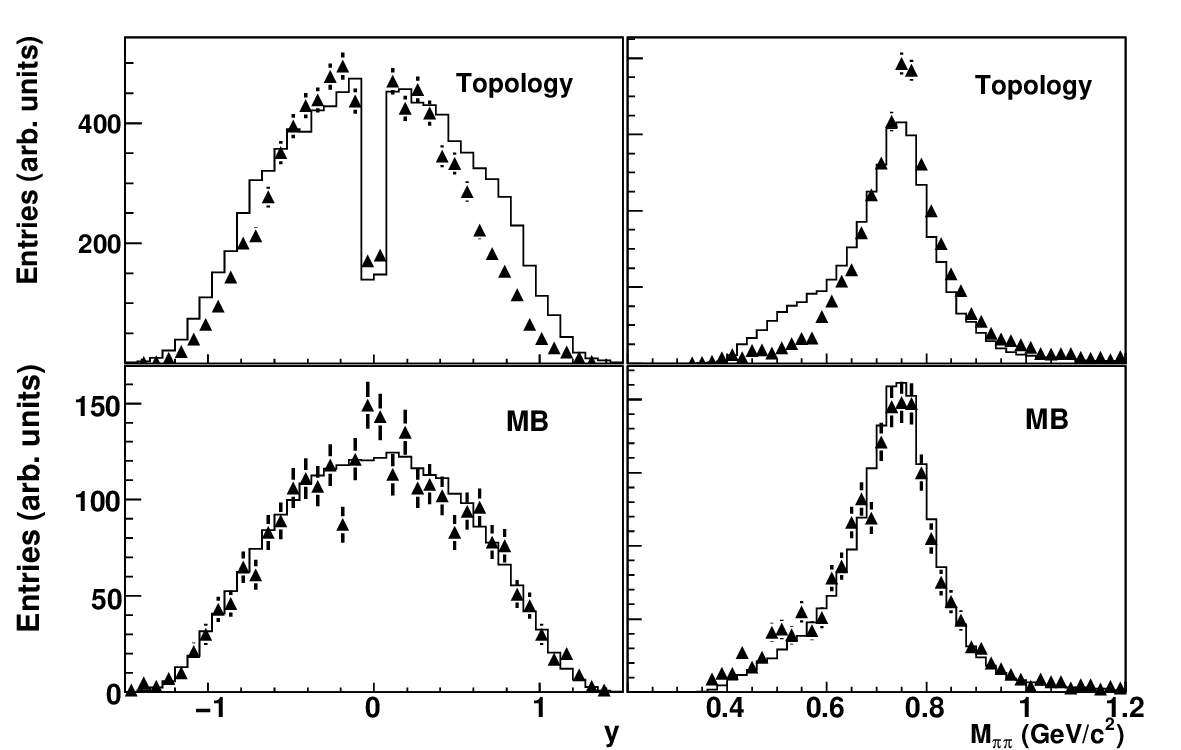}
\caption[]{Rapidity (left) and $M_{\pi\pi}$ (right) of the $\pi^+\pi^-$ 
distributions for the topology (exclusive $\rho^0$, top)
and MB (Coulomb breakup, bottom) samples.  The
points with statistical error bars are the data, and the histograms
are the simulations. The 'notch' in the topology data around $y=0$ is
due to the explicit rapidity cut to remove cosmic-ray backgrounds.
\label{fig:checks}}
\end{figure}

To understand the effect of detector resolution,  $\rho^0$ events were generated following the
Klein-Nystrand distributions, and passed through the detector simulation and reconstruction.
Figure \ref{fig:checks} compares the
rapidity and $M_{\pi\pi}$ distributions of the data and simulations.
The agreement is good for the minimum bias data, less so for the 
topology data.    This is most likely due to an imperfect topology trigger simulation;
the effect of this is treated as a systematic error.   
  
The applied cuts select both directly produced $\pi^+\pi^-$ pairs \cite{directpipi}
and $\rho^0$.  Direct $\pi^+\pi^-$ and $\rho^0$ decays are
indistiguishable, so the two processes interfere.  
The $\rho^0$ mass peak and direct pion fraction are
consistent with earlier gold-gold photoproduction studies \cite{STARrho^0}.
The two subchannels should have the same spin/parity and quantum mechanical behavior, 
so we do not distinguish between  them.  If $\pi^+\pi^-$ pairs are produced by
a different production mechanism, such as Pomeron-Odderon interactions \cite{Odderon}, then this 
assumption might not hold.
 
\begin{figure}
\label{tslopetopo}
\includegraphics[scale=0.38]{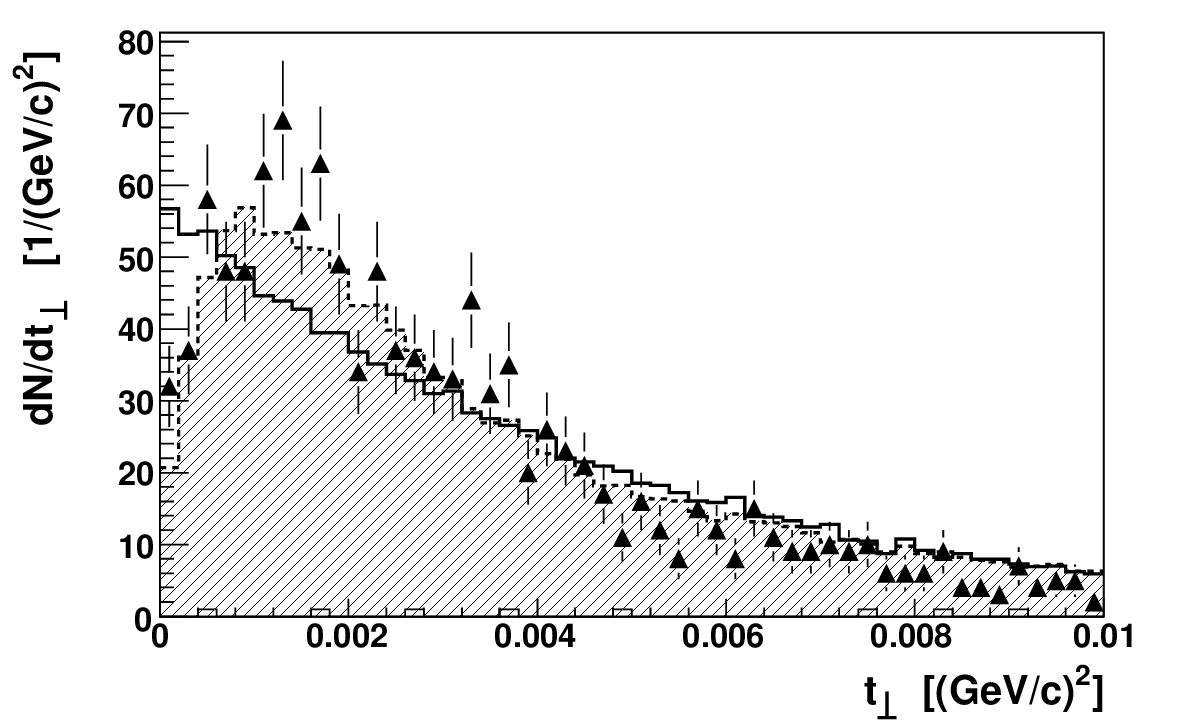}
\caption[]{Raw (uncorrected) $\rho^0$  $t_\perp$-spectrum 
in the range $0.0 < |y|<0.5$ for the MB data.  The points are data,
with statistical errors. 
The dashed (filled) histogram is a simulation with an interference term (``Int"),
while the solid histogram is a simulation without interference (``NoInt").
The handful of events histogrammed at the bottom of
the plot are the wrong-sign 
($\pi^+\pi^+ + \pi^-\pi^-$) events, used to estimate the combinatorial
background.
\label{fig:uncorr}}
\end{figure}

Figure \ref{fig:uncorr} compares the uncorrected $|\eta|<0.5$  MB data 
and simulations based on Refs. \cite{usPRC, interfere, baltzus} with and without interference, as a function of  $t_\perp = p_T^2$.  
At RHIC energies, the longitudinal component of the 4-momentum transfer
is small, so $t_\perp \approx t$.  
The measured $dN/dt$ spectrum is roughly exponential, but with 
a significant downturn for $t_\perp<0.0015$
GeV$^2$, consistent with the predicted interference. 
The no-interference histogram is almost exponential, 
$dN/dt \propto \exp{(-kt_\perp)}$, where $k$ is related to the nuclear radius 
\cite{interfere,ting}, even though the Klein-Nystrand calculation uses a Woods-Saxon distribution
for the gold density.  The Hencken-Baur-Trautmann calculation uses a Gaussian distribution for the nuclear
density, but is also fairly well fit by an exponential. 
The interference in different $y$
ranges is determined using a Monte Carlo simulation which follows
the Klein-Nystrand calculations. 

\begin{figure}[hbtp]
\includegraphics[scale=0.43]{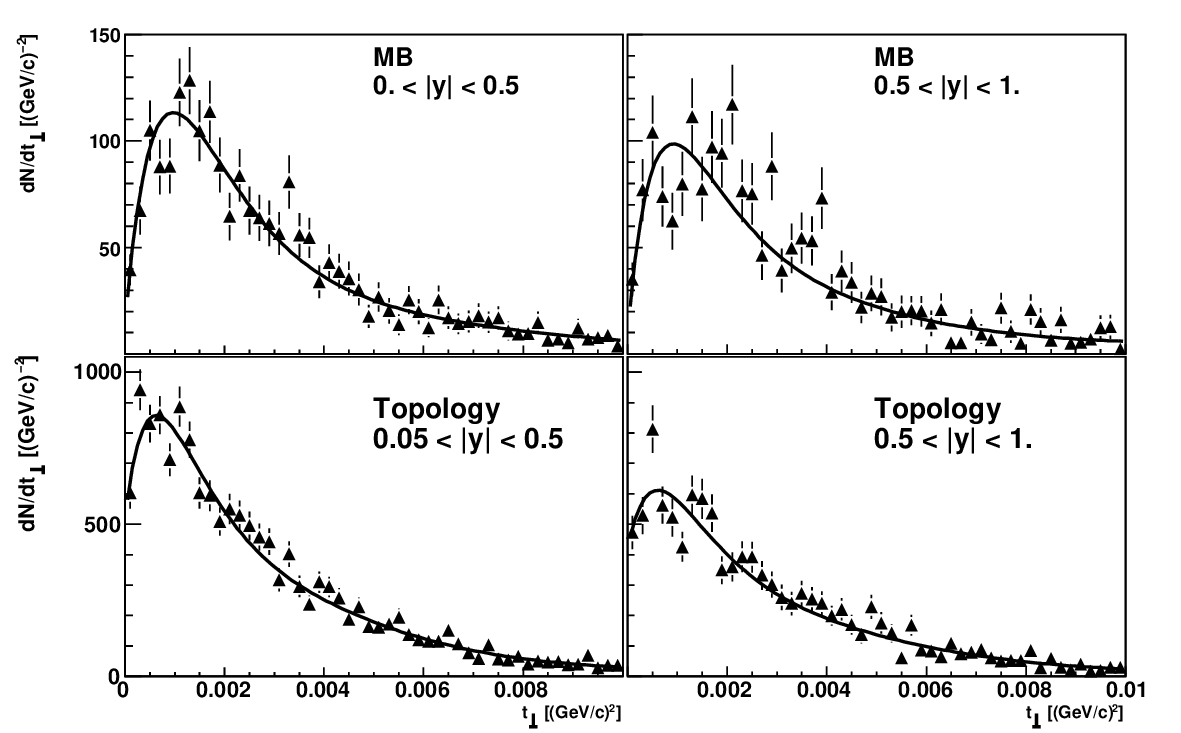}

\caption[]{Efficiency corrected $t_\perp$ spectrum for $\rho^0$ from
(top) minium bias and (bottom) topology data, for mid-rapidity (left) 
and larger rapidity (right) samples.   The points are the
data, while the solid lines are the results of fits to Eq. (\ref{eq:fit}).}
\label{fig:data}
\end{figure}

Figure \ref{fig:data} shows the
efficiency corrected MB and topology data.  All four panels show a dip as
$t_\perp\rightarrow 0$.  As expected, this dip is broader for the MB data because
$\langle b\rangle$ is smaller. The suppression at $t_\perp=0$  is larger for the 
small-rapidity samples
because the amplitudes for the two photon directions are more similar.
The efficiency is almost independent of $p_T$, so Fig. \ref{fig:uncorr} is not
very different from the efficiency corrected $t_\perp$ spectra 
in Fig. \ref{fig:data}. The main effect of the detector response is 
$p_T$ smearing due to the finite momentum resolution.

The $dN/dt$ spectrum is fit by the 3-parameter form
\begin{equation}
{dN\over dt} = A \exp(-kt) [1+ c(R(t)-1)],
\label{eq:fit}
\end{equation}
where 
\begin{equation}
R(t_\perp) = {\text{Int}(t_\perp)\over \text{NoInt}(t_\perp)}
\end{equation}
is the ratio of the simulated $t_\perp$-spectra with and
without interference. 
For $t_\perp \gg 0.01 {\rm GeV}^2$, 
$R(t_\perp)\rightarrow 1$, but for $t_\perp \le 0.01$ GeV$^2$, $R(t_\perp)\ne 1$.  
$A$ is the overall (arbitrary) normalization and $c$ gives the degree of
spectral modification; $c=0$ corresponds to no interference, while
$c=1$ is the predicted Klein-Nystrand interference.  
Table 1 gives the fit results.

$R(t_\perp)$ was determined using a simulation that includes the
detector response, and then fit to
two analytic functions:   $R(t_\perp) = \Sigma_{i=0}^n a_i/(t_\perp+0.012{\rm GeV}^2)^i$ and
$R(t_\perp) = \Sigma_{i=0}^n a_it_\perp^i$. Our results use the first polynomial with $n=5$; the 
second polynomial and different values of $n$ were used to estimate the 
fitting uncertainties. 

\begin{table}
\begin{ruledtabular}
\begin{tabular}{lcccc}
Dataset & $A$  & $k$ & $c$ & $\chi^2$/ \\
 &  & (GeV$^{-2}$) &   & $DOF$ \\
\hline
MB, $|y|< 0.5$& $6,471\!\pm\!301$ & 
$299\pm12$  & $0.92\pm 0.07$ & 45/47\\
MB, $0.5\!<\!|y|\!<\!1.0 $& $5,605\!\pm\!330$ & 
$303\pm15$ & $0.92\pm 0.09$ & 76/47 \\
T, $0.05\!<\!|y|\!<0.5$ & $11,070\!\pm\!311$   &
$350\pm8$ & $0.73\pm 0.10$ & 53/47 \\
T, $0.5 <|y|< 1.0$  & $12,060\!\pm\!471$  & 
$333\pm11$ & $0.77\pm 0.18$  & 64/47 \\
\end{tabular}
\end{ruledtabular}
\caption[]{The results of fitting Eq. \ref{eq:fit}
to the four data sets.  Here, T is for topology. The $\chi^2/DOF$ are
discussed in the text.}
\label{t:fitout}
\end{table}

The weighted average of the four $c$ values is $c = 0.84 \pm 0.05$.  
The $k$ values for the MB and topology datasets differ by 15\%.  This
may be due to the different $b$ distributions.  The photon flux
scales as $1/b^2$, so the photon flux on the `near'
side of the nucleus is larger than on the `far' side.  As $b$ decreases, 
$\rho^0$ production is increasingly concentrated on the near side, and
the apparent production volume drops, reducing $k$. A calculation with different assumptions 
may  predict a different electric field variation which leads to a smaller difference in $k$.

Two of the fits have $\chi^2/DOF$ significantly larger than 1.  
The $\chi^2$ did not decrease with
different fit functions for $R(t)$, variations 
of the nuclear radius in the interference calculations, background level, or modifications
to the detector simulation.  
When the $\chi^2/DOF>1$, we scale up the fit errors on $c$ by $\sqrt{\chi^2/DOF}$;
this excess error may  have theoretical and/or experimental origin.  With the
scaled errors, the weighted average is $c=0.86\pm0.05$.

Systematic errors come from instrumental effects, background, 
fitting, and theoretical issues. The major instrumental effects were due
to the topology trigger;  we apply a 10\%
systematic error to the topology data to account for this.

This analysis is sensitive to any $\rho^0$ $p_T$-dependent
efficiency variation.  The decay pions have a typical 
$p_T$ of about 300 MeV/$c$, where the detection efficiency is high 
and almost $p_T$-independent \cite{TPC}.  However, the $\rho^0$ $p_T$ resolution,
about 7.5 MeV/$c$, smears the $t_\perp$ spectrum in the two lowest $t_\perp$ bins.  
To study detector effects, 
we fit the raw (uncorrected) $t_\perp$ spectrum with the raw Monte Carlo output; 
this reduced $c$ by 18\% \cite{QM2004}, mostly due to the $p_T$ smearing.
We assume conservatively that the detector simulation
is only 80\% effective, and assign a 4\% systematic error on $c$
to account for non-trigger detector effects. 

Backgrounds were estimated by including
like-sign pairs ($\pi^+\pi^+ + \pi^-\pi^-$) in the fits.  $c$ 
changed by less than 0.5\%. We assign a 1\% systematic error due to backgrounds. 

The uncertainty due to fitting was evaluated by comparing
results using the two different polynomial forms of 
$R(t)$ for both $n=4$ and $n=5$; $c$ varied by an average of 1\%. 
The effect of an imperfect form factor model 
was studied by varying the nuclear radius in the simulations.
A $\pm$ 20\% change in nuclear radius changed $c$ by 3\%.  
We assign a 4\% systematic error due to the fitting procedure. 

The theoretical uncertainties are difficult to
evaluate. Our simulation follows Refs. \cite{interfere, baltzus}
in detail, but those calculations themselves contain uncertainties.
The two theoretical models agree well for exclusive $\rho^0$ production.
For  $\rho^0$ production accompanied by mutual Coulomb excitation, there is 
some disagreement, but the Klein-Nystrand model has a more detailed excitation calculation, and
so may be more accurate. We assign a 5\% systematic error due to theoretical issues. 

Combining these systematic errors in quadrature results in an 8\% (13\%) systematic
error for the MB (topology) data.  Adding
the four results in quadrature, including the systematic errors, leads to an 
interference that is $87 \pm 5 ({\rm stat.}) \pm 8\%$(syst.) of the expected level.  

Because $\rho^0$s decay so rapidly, $\gamma\beta c\tau\ll \langle
b\rangle$, the $\rho^0$ decay points are well separated in space-time,
and the two amplitudes cannot overlap and interfere until after the
decay occurs.  The interference must involve the $\pi\pi$ final states \cite{physlett}.  This
interference is only possible if the post-decay $\pi\pi$ wave functions 
retain amplitudes for all possible $\rho^0$ decays, at least until the wave functions from
the two ion sources overlap.  The $\pi^+\pi^-$  wave function is not factorizable and is thus an
example of the Einstein-Podolsky-Rosen paradox \cite{EPR}.  Unlike previous tests
of non-locality, the interference involves continuous variables, momentum and position
\cite{physlett}.  

In conclusion, we have measured the interference between $\rho^0$ production at two
sources (the two nuclei) by observing the $\pi^+\pi^-$ decay products. 
We observe the interference at $87\pm5$(stat.)$\pm8$ (syst.)\% of the expected level. 
This shows that the final state wave function retains amplitudes for all possible decays, 
long after the decay occurs.
The maximum decoherence (loss of interference) is less than 23\% at the 90\% confidence
level.  

We thank the RHIC Operations Group and RCF at BNL, and the NERSC Center 
at LBNL and the resources provided by the Open Science Grid consortium 
for their support. This work was supported in part by the Offices of NP 
and HEP within the U.S. DOE Office of Science, the U.S. NSF, the Sloan 
Foundation, the DFG Excellence Cluster EXC153 of Germany, CNRS/IN2P3, 
RA, RPL, and EMN of France, STFC and EPSRC of the United Kingdom, FAPESP 
of Brazil, the Russian Ministry of Sci. and Tech., the NNSFC, CAS, MoST, 
and MoE of China, IRP and GA of the Czech Republic, FOM of the 
Netherlands, DAE, DST, and CSIR of the Government of India, Swiss NSF, 
the Polish State Committee for Scientific Research,  and the Korea Sci. 
\& Eng. Foundation.

\vfill\eject

\begin{references}
\def\etal{{\it et al.}}

\bibitem{reviews}C. A. Bertulani, S. R. Klein and J. Nystrand,
Ann. Rev. Nucl. Part. Sci. {\bf 55}, 271 (2005);
G. Baur {\it et al.}, Phys. Rep. {\bf 364}, 359
(2002); F. Krauss, M. Greiner and G. Soff, Prog. Part. Nucl.
Phys. {\bf 39}, 503 (1997); K. Hencken {\it et al.}, 
Phys. Rept. {\bf 458}, 1 (2008). 

\bibitem{usPRC}S. Klein and J. Nystrand, Phys. Rev. {\bf C60}, 014903
(1999).

\bibitem{STARrho^0}C. Adler {\it et al.}, Phys. Rev. Lett. {\bf 89},
272302 (2002); B. I. Abelev {\it et al.}, Phys. Rev. {\bf C77}, 034910 (2008).

\bibitem{strikman}L. Frankfurt, M. Strikman and M. Zhalov,
Phys. Rev. {\bf C67}, 034901 (2003); 
L. Frankfurt, M. Strikman and M. Zhalov, Phys. Lett. {\bf B537}, 51
(2002).

\bibitem{baltzus}A. Baltz, S. Klein and J. Nystrand,
Phys. Rev. Lett. {\bf 89}, 012301 (2002).

\bibitem{interfere}S. Klein and J. Nystrand, Phys. Rev. Lett.
{\bf 84}, 2330 (2000).

\bibitem{BHT}K. Hencken, G. Baur and D. Trautmann, Phys. Rev. Lett.
{\bf 96}, 012303 (2006).

\bibitem{phase}T. H. Bauer {\it et al.}, Rev. Mod. Phys.  {\bf 50}, 261
(1978).

\bibitem{physlett}S. Klein and J. Nystrand, Phys. Lett. {\bf A308},
323 (2003).  

\bibitem{EPR}A. Einstein, B. Podolsky and N. Rosen, Phys. Rev.
{\bf 47}, 777 (1935).

\bibitem{decoherencereview}M. Schlosshauer, Rev. Mod. Phys. 
{\bf 76}, 1267 (2005). 

\bibitem{factorize}G. Baur {\it et al.}, Nucl. Phys. {\bf A729}, 787
(2003). 

\bibitem{bcw}A. J. Baltz, C. Chasman and S. N. White, Nucl. Instrum
\& Meth. {\bf A 417}, 1 (1998). 

\bibitem{newbaltz}A. J. Baltz, Phys. Rev. Lett. {\bf 100}, 062302 (2008).

\bibitem{TPC}M. Anderson \etal, Nucl. Instrum \& Meth. {\bf B499}, 659
(2003); M. Anderson \etal, Nucl. Instrum \& Meth. {\bf B499}, 679
(2003).

\bibitem{trigger}F. S. Bieser {\it et al.}, Nucl. Instrum \&
Meth. {\bf B499}, 766 (2003).

\bibitem{ZDCs}C. Adler {\it et al.}, Nucl. Instrum. \& Meth.  {\bf
A470}, 488 (2001).

\bibitem{directpipi}P. S\"oding, Phys. Lett. {\bf 19}, 702 (1966).

\bibitem{Odderon}L. Motyka, preprint arXiv:0808.2216

\bibitem{ting}G. McClellan {\it et al.}, Phys. Rev. {\bf D4}, 2683 (1971).

\bibitem{QM2004}S. Klein for the STAR Collaboration, nucl-ex/0402007.

\end{references}
\end{document}